\documentstyle[aps,prl,floats,twocolumn]{revtex}
\begin{document}
\draft
\twocolumn[\hsize\textwidth\columnwidth\hsize\csname @twocolumnfalse\endcsname
\title{Scaling theory of two-dimensional metal-insulator transitions}
\author{V. Dobrosavljevi\'{c}$^1$, Elihu Abrahams$^{1,2}$, E. Miranda$^1$, and
Sudip Chakravarty$^3$}
\address{$^1$National High Magnetic Field Laboratory, Florida State
University\\1800 E. Paul Dirac Dr., Tallahassee, Florida 32306\\
$^2$Serin Physics Laboratory, Rutgers University,
Piscataway, NJ 08855-0849\\
$^3$Department of Physics and Astronomy, University of California Los Angeles\\
Los Angeles, CA 90095-1547}

\date{\today}
\maketitle

\begin{abstract}

We discuss the recently discovered two-dimensional metal-insulator
transition in zero magnetic field in the light of the scaling theory
of localization.  We demonstrate that the observed symmetry
relating conductivity and resistivity follows directly from the quantum
critical
behavior associated with such a transition. In addition, we show that very
general
scaling considerations imply that any disordered two dimensional metal is a
{\em perfect metal},
but most likely {\em  not} a Fermi liquid.
\end{abstract}

\pacs{71.30.+h,72.15.Rn,72.10.Bg}

]
In an influential 1979 paper \cite{g4}, simple
scaling  arguments were put forward which have shaped much of our thinking
about metal-insulator transitions (MITs) in disordered systems. Focusing on
noninteracting electrons, the authors demonstrated that in two
dimensions (2D) even weak disorder is sufficient to localize the electrons,
and no true metallic behavior is possible at $T=0$. Ever since,
it has been widely believed that two is the lower critical dimension for
MITs in disordered systems and this prompted much theoretical
activity based on expansions around dimension $d=2$. The scaling ideas were
quickly
extended to interacting systems, but despite many years of painstaking
effort, no satisfying theoretical picture has emerged for the interacting case.
Nevertheless, the belief that all the states are localized in $d=2$ has
remained largely unquestioned.

Recently, a remarkable experiment \cite{krav1} performed on a 2D
electron gas in zero magnetic field suggested that the conventional
picture might be incomplete. In this work, fairly convincing evidence
was presented which supports the existence of a true MIT in 2D. Despite the
beauty
and elegance of the data, this  work has met considerable resistance, largely
due to its
apparent  conflict with the scaling theory of localization as well as its
uniqueness.

The major assumption of \cite{g4} was based on an earlier idea of
Thouless \cite{djt} about the length (scale) dependence of the conductance.
In \cite{g4} it is  asserted that the ``beta function''
$\beta (g)=d [\log(g)]/d[\log (L)]$ is a function of the conductance $g$
itself,
but not an explicit function of the length scale $L$. Now, $\beta (g)$ is known
in the two limits of very large and very small disorder, and it is reasonable
to assume that it is continuous (smooth) in between. From Ohm's law,
$\beta (g)= (d-2)$ for $g \rightarrow +\infty$, while for $g$ small, one
expects exponentially localized states, so that $\beta (g)\rightarrow
 -\infty$ in this limit. Since the metallic behavior is possible only
for $\beta (g) > 0$, the form of $\beta (g)$ at large $g$ is sufficient to
determine the stability of the metallic phase. In particular, for
noninteracting electrons, this has a form $\beta (g)= (d-2)-1/g+...$
\cite{loop},
indicating that the metallic phase is unstable in $d \le 2$.

However, it is important to emphasize that the last step in this
analysis is valid {\em only} for noninteracting electrons; the
scaling theory is only complete in the absence of electron-electron
interaction. In
the  following, we show that the existence of a 2D transition does not
contradict any
general scaling principles. However, we argue that within the scaling
theory such a transition has a number of unusual
features, many of which are already apparent in the existing data.

No fundamental principle
requires that $\beta (g)$ be  monotonic, or in fact negative in 2D when $g$ is
large.
Instead,  if we assume that the
leading behavior \cite{analyt} at large $g$ is
\begin{equation}
\beta (g)= (d-2)+A/g^{\alpha}+...,\  A>0,
\end{equation}
then it is
clear that the beta function for all $d \geq 2$ has to change sign at some
finite
$g=g_c$, leading, in particular, to a metal-insulator transition in
2D.
Alternatively, the
existence of a MIT in 2D {\em requires} that $\beta (g) > 0$ at $g$ large,
where the exponent $\alpha$ simply parameterizes how the system approaches the
metallic limit as
$g\rightarrow\infty$.

This idea is not new. Such a scenario is realized for
non-interacting electrons in the presence of spin-orbit scattering \cite{so}.
In this case, $\alpha=1$ and a MIT is expected in 2D.
However, this behavior has never been taken too seriously
in the context of real 2D systems  because
it is known that adding  interactions to the spin-orbit
universality class results in {\em reversing} the sign of the
first quantum correction \cite{loop}, and all the states again end up localized
in
2D. A more complicated situation was found in the absence of spin-orbit and
time-reversal breaking perturbations.
There, an interaction driven enhancement of the conductivity
{\em was} found at weak disorder which was expected
to overwhelm the localizing effects of coherent backscattering
(weak localization) \cite{fink}. However, the analysis
revealed that the effective interaction strength diverges upon
scaling, making it difficult to determine what will actually happen
at long scales or low temperatures \cite{fink}. Nevertheless, from what is
known it appears that the relevant quantum corrections due to interactions can
only
{\em enhance} the conductivity at low temperatures, so that the
possibility of a 2D metal remains open.

At present, the abovementioned problems
in the interaction case make it difficult to determine the form
of the leading quantum corrections even at weak disorder. Rather
than elaborating on this, we shall {\em assume}
that interactions lead to a positive $\beta$ at $g$ large, so that the 2D
metal exists. The first question that one can address is what is
the low temperature form of the conductivity in the metallic phase?
Within the scaling theory this is determined
{\em only} by the form of $\beta (g)$ at $g$ large. Assuming a form of
Eq.\ (1), and integrating the beta function from $\ell$ (mean free path)
to $L$ (sample size), we find
\begin{equation}
g(L)=[g_0^{\alpha} +A\alpha \log (L/\ell )]^{1/\alpha}.
\end{equation}
If we consider an infinite sample at finite temperature, then
$L_{eff}\sim T^{-1/z}$. Here $z$ is the dynamical exponent
associated with the {\em metallic} ($g=\infty$) fixed point. One
expects $z=2$ in a conventional diffusive regime, in the
absence of dangerously irrelevant variables \cite{note}.   We thus
find that at low temperatures in the metallic phase
\begin{equation}
g(T)\sim \log^{1/\alpha} (T_0/T).
\end{equation}
In other words, the conductance will {\em diverge},
i.e. the resistance will {\em vanish} at $T=0$ in the metallic phase,
albeit in a weak, logarithmic fashion. This should be true at
sufficiently
low temperatures throughout the metallic phase, i.e. the temperature
dependence should have a {\em universal} form,  as in Fermi liquid
theory. However, we emphasize that this is {\em not} a Fermi liquid, not
only because of its unusual temperature dependence, but more
fundamentally,
since we expect the nature of these electronic states to change completely
if the interactions are turned off: they would localize.

Another experimentally relevant question is what is the temperature
dependence of the conductance in the {\em quantum critical region}
associated with the metal-insulator transition. Within scaling theory,
this question can be answered very precisely, as follows. In our
formulation, the transition occurs at $g=g_c$ where the
beta function changes sign, i.e. $\beta (g_c )=0$.
%
%
We recall that deep in the insulating phase, the
beta function is {\em logarithmic} in $g$ \cite{g4}. It is thus
plausible to introduce the quantity $t=\log (g/g_c )$ as a natural
scaling variable. To determine the
critical behavior it is sufficient to consider the linear
approximation to the beta function near its zero at $t=0$. The slope of
$\beta(t)$ at
$t=0$ determines the correlation length exponent $\nu$:
\begin{equation}
\beta (t)=\frac{dt}{d(\log L)}\approx \frac{1}{\nu} t + {\rm O}(t^2)
\end{equation}
By integrating this equation from $\ell$ to $L$,
we find
\begin{equation}
t(L) =t_0\left( \frac{L}{\ell}\right)^{\frac{1}{\nu}},
\end{equation}
where $t_0 = \log (g_0/g_c)$ is determined by the starting value $g_0$ of the
conductance at scale $\ell$. In the critical region, where we start with a
$g_0$ very close to $g_c$, $t_0\approx (g_0-g_c)/g_c \propto \delta n$,
where $n$ is the carrier concentration, and
$\delta n =(n-n_c )/n_c $ measures the distance
to the critical point. Then from Eq.\ (5), we have for the conductance
\begin{equation}
g(L)=g_c \exp\left[ A \delta n
( L/\ell)^{\frac{1}{\nu}}\right].
\end{equation}

At non-zero temperature, the length scale $L$ is determined by the
temperature through the dynamical exponent as $T\sim L^{-z}$ and we obtain the
temperature dependence of the conductance
\begin{equation}
g(\delta n,T)=g_c \exp \left( {\rm sgn}(\delta n) A
[T_0(\delta n)/T]^{\frac{1}{\nu z}}\right),
\end{equation}
where we have defined a crossover temperature $T_0$ corresponding to the
inverse correlation time as
\begin{equation}
T_0 (\delta n)\sim |\delta n| ^{\nu z},
\end{equation}
Here, $A$ is an unknown dimensionless constant of order one.
Let us define the scaled conductance
as $g^* (\delta n,T) =
g(\delta n,T)/g_c$. From Eq. (7), we then immediately find a striking
symmetry
relating the conductance on the metallic side ($\delta n > 0$)
to the resistance
on the insulating side ($\delta n < 0$) of the transition at $g^* = 1$ (i.e. at
$\delta
n = 0$):
\begin{equation}
g^* (\delta n, T)= 1/g^* (-\delta n, T).
\end{equation}
Remarkably, precisely this behavior is clearly seen in the
experiments~\cite{krav1}, as emphasized in Ref.~\cite{krav3}.

In related work \cite{krav2}, a similar symmetry for the
dependence on electric field was found for the same 2D system. If one
uses the conventional electric field scaling hypothesis \cite{E}, the
preceding arguments carry over with the electric field $E$ replacing $T$
and $1+z$ replacing $z$, so that this symmetry too appears as a
consequence of the scaling argument we have presented. In fact, the
behavior holds whatever the exponent relating the electric field to a
characteristic length.

In order to fully appreciate the  significance of these findings, we should
carefully qualify them and comment on the range of their validity.

(1) In \cite{krav3}, the authors point out a similarity of this symmetry
with the one observed in the context of the quantum Hall
liquid-to-insulator (QHI) transitions \cite{qhi}. However, we emphasize
that the theory described in \cite{qhi} is fundamentally different from
that of the present paper. The symmetry found in a 2D MIT is restricted
to the {\em quantum critical region}, which is defined by the crossover
temperature scale, i.e. it is expected to hold only for $T > T_0 (\delta
n)$.  In fact, the experiments completely confirm this expectation, as
the authors themselves point out~\cite{krav3}. A careful examination of
the data in \cite{krav1} reveals that the temperature at which a
departure from symmetry is observed is $T\approx T_0 (\delta n)$, with
$T_0$ shown in the inset of Fig.\ 3 of \cite{krav3}. The same feature is
clearly seen also in the earlier data of \cite{krav1}, where in Fig.\ 4,
the resistance is plotted as a function of the reduced temperature $\tau
= T/T_0 (\delta n)$. Here, the obvious symmetry of the resistance and
the conductance on the respective metallic and insulating sides of the
MIT is seen and it is violated below $T\approx T_0 (\delta n)$, as
expected. The duality observed in QHI is thought to be a charge-flux
duality that holds for the full Chern-Simons Lagrangian, ignoring the
roles of irrelevant operators, cutoffs, or disorder. If this is so, it
has little resemblance to the symmetry in MIT. On the other hand, it may
be that the data in QHI transition can be understood from the present
perspective, where $(B-B_c)\propto (1/\nu-1/\nu_c)$ is the control
parameter that tunes the system through a quantum critical point; here
$B$ is the magnetic field, and $\nu$ is the filling fraction.

(2) We emphasize that the above results were derived by
{\em linearizing} the beta function around the transition
point $g=g_c$. The symmetry is then simply a consequence
of the smoothness of the beta function in
the critical region. However, without further knowledge about the
form of the beta function, the result is strictly speaking
valid only for $\delta n \ll 1$, i.e. it only defines the
form of the  leading high-temperature correction
in the quantum critical regime. In other words, to be consistent,
one should keep only the leading term of the expansion of the exponential of
Eq.\ (7):
\begin{equation}
g(\delta n, T)\approx g_c \left[1+ A\delta n /T^{(1/\nu z)}\right].
\end{equation}
This expression is generally expected to be valid only for
$|g(T)-g_c|/g_c \ll 1$. On the other hand,  if we {\em assume} that the
linearized expression  for the beta function [Eq.\ (4)] is a good
approximation over an appreciable conductance range
$g_{min} < g_c < g_{max}$, then the full  exponential
behavior is valid in that range.
The experimental data reveal that the symmetry
is found in a much broader range than allowed by the
form of Eq.\ (10). In fact, the data clearly display an
exponential temperature dependence as in Eq.\ (7), which is particularly
striking in the metallic phase, where the resistance is found
to drop almost by an order of magnitude  within the
``symmetry" regime. The deviations from leading behavior
are more clearly seen in Fig.\ (2b) of \cite{krav3},
where the conductance and as well as the resistance are plotted at
a fixed temperature, for a range of carrier concentrations.
The two curves show perfect mirror symmetry which holds over
an extended range of concentrations, where the conductance shows
a dramatic exponential dependence on concentration,
in agreement with Eq.\ (7). This should be contrasted with the
leading correction given by Eq.\ (10), where the density
dependence is {\em linear}, as expected only for $\delta n \ll 1$.

(3) We conclude that the experimental results of
\cite{krav1,krav2,krav3} go beyond just confirming the expectations
based on general scaling arguments in the quantum critical region. They
also provide striking evidence about the form of the beta function in
the critical region. In particular, they indicate that in a wide range
of conductances $1/4 < g/g_c < 4$ the beta function is well-approximated
by the linear expression in $t=\log (g/g_c )$ as in Eq.\ (4).  How can
we rationalize this finding? Deep in the insulating regime, ($g \ll
g_c$) the beta function is {\em exactly} given by $\beta (g) \sim
\log(g)$.  The above experimental result can thus be interpreted as
evidence that the same slow logarithmic form of the beta function
persists beyond the insulating limit well into the critical
regime. Further support for this idea can be found in the experimental
findings of Hsu and Valles \cite{valles} on ultrathin films. While no
MIT was found, their data yields a beta function which is close to
linear in $\log(g)$ (``strong localization") all the way to the standard
crossover conductance $g \simeq e^2/2h$. If it is generally true, this
feature could be used as a basis of approximate calculations of the
critical exponents. In contrast to the well-known $2+\varepsilon$
expansion, here one would try to obtain the form of the beta function in
the critical region by an expansion around the {\em strong disorder}
limit.



(4) There are physical reasons which support our conjecture that the
beta function must be a logarithmic function close to $g_c$, if $g_c$ happens
to fall in the regime of strong disorder, which the experiments discussed here
indicate.
The assumption of the scaling theory is that the conductance $g$ is the
only important parameter at large distances and low energies. That is, the
distribution
of conductances of a hypercube of linear dimension $L$ is so sharply peaked
that the most
probable value is the only value, hence the mean value. This assumption can be
justified in the limit of weak disorder, where the sample to sample
fluctuations of the
physical properties are weak, but not when the disorder is strong and the
distribution of
the conductances is broad. At strong disorder, it is more plausible to assume,
if we still want to  describe the problem with a single scaling variable,
that the distribution is sharply peaked when described in terms of $\log g$
\cite{g42},
or more generally in terms of a function $\log[\phi(g)]$, where
$\phi(g)$ is a smooth function of $g$ that tends to $b_{-1}/ g + b_0 + b_1 g +
\ldots$,  as $g\to 0$, but tends to $a_0 + a_1/ g + a_2/g^2\ldots$, as $g\to
\infty$.
At strong disorder, it is natural to believe that the
Green's function between two points can be computed in a hopping
parameter expansion\cite{pwa58}.
It is plausible that the impurity averaged square of the Green's function for
$p$ hops on an effective lattice of spacing $a$ larger than the mean free path
is
\begin{equation}
\overline{|G^2(p)|}=\left[{\phi(g_c)\over \phi(g)}\right]^p.
\end{equation}
If $g>g_c$, the perturbation theory does not converge as $p\to
\infty$; this signals a failure of the localization assumption. When $g<g_c$,
this is not so, and we can define a localization length. However, for a finite
system of
linear dimension $L$, one can meaningfully use this equation for all
$g$, because $p$ is restricted to be less than $(L/a)$. This allows us to
derive
the beta function, as follows. From Eq.\ (11),
the physical localization length is
\begin{equation}
\xi={2a\over  \log\left[\phi(g)/\phi(g_c)\right]}.
\end{equation}
If we keep the physical $\xi$ fixed as we vary the lattice spacing, that is,
set
$d\xi /da=0$, we get
\begin{equation}
\beta = {\partial  \log \phi(g)\over \partial \log a}= \log \left[{\phi(g)\over
\phi(g_c)}\right].
\end{equation}
To check that this is correct, consider the limit $g, g_c \to 0$. We find that
\begin{equation}
{\partial  \log g\over \partial \log a}=(1+{b_0\over
b_{-1}}g+\cdots) \log\left({g\over g_c}\right),
\end{equation}
which is precisely the locator expansion \cite{g4,pwa58}.
The difficult task is to be able
to compute
$\phi(g)$ from a microscopic model.  Note, however, that local
electron-electron
interactions should not pose any special problems in the locator expansion.

(5) The symmetry associated with the quantum critical
region is specific to 2D systems. In particular, standard
arguments suggest that in general dimension, the conductivity as a function of
reduced concentration and temperature should assume the scaling form
\begin{equation}
\sigma(\delta n, T)\sim T^{(d-2)/z}f(\delta n/ T^{1/\nu z}).
\end{equation}
Here, $f(x)$ is a universal scaling function such that $f(0)=1$, in
the absence of dangerously irrelevant variables. As a result of the extra
temperature prefactor present when $d>2$, the conductivity will vanish
at the  transition ($\delta n=0$, $T=0$), ruining the reflection symmetry
of the
quantum critical region. This fact would make it very difficult,
if not impossible, to extract the information about the details of the beta
function from the temperature dependence in the quantum critical region when
$d>2$.

(6) The samples showing the 2D MIT are distinguished from those in
previous work, as pointed out by the authors of \cite{krav1}, by the
fact that the density $n$ is so low that the Coulomb interaction $U
\propto \sqrt{n}$ is immense, almost an order of magnitude greater than
the Fermi energy $E_F \propto n$.  Furthermore, for small $n$, high
mobility $\mu$ is required to reach the critical conductance $g_c = n_c
e \mu \simeq e^2/3h$. This also leads to the conclusion that $k_F l
\lesssim 1$ for the samples showing MIT. Thus it is misleading to assume
that because the mobilities are high they are weakly disordered .  These
conditions both lead to the conclusion that the one-electron picture of
\cite{g4} cannot be expected to be valid and may explain why the MIT is
not observed in some samples.

In summary, we have discussed the recently discovered 2D
metal-insulator transition in zero magnetic field in the light of the
scaling scenario for localization.  We have shown that very general scaling
considerations imply that any disordered 2D metal is a {\em
perfect metal}, but most likely {\em not} a Fermi liquid. In addition,
we have  demonstrated that the observed symmetry relating conductivity and
resistivity follows directly from the quantum critical behavior
associated with such a transition. Furthermore, the fact that this symmetry is
found over
an extended range of conductances provides important information  about the
form of the
beta function in the critical region.

The support of NSF grants DMR 96-32294
(EA) and DMR 95-31575 (SC) is acknowledged. EA thanks the NHMFL for its
hospitality.
VD was supported by the NHMFL/FSU and the Alfred P. Sloan Foundation. EM
was supported by the NHMFL/FSU.

We thank P.W. Anderson, R. Bhatt, N. Bonesteel, L. Engel, S. Kravchenko, C.
Nayak, D. Popovic, M. Sarachik, D. Simonian, and X-G. Wen
for useful discussions.


\begin{references}
\bibitem{analyt} If the beta function is to be analytic, then $\alpha$ is
restricted
to integer values.
\bibitem{note} Note that there can be
another $z$ associated with the critical region of the MIT.
\end{references}
\end{document}